\documentclass[nofootinbib,amsmath,twocolumn,notitlepage,preprintnumbers,longbibliography]{revtex4-1}
\usepackage{multirow}
\usepackage{amssymb,esvect,amsmath,graphicx,latexsym,amsthm,slashed,eso-pic}
\usepackage{xspace}
\usepackage[colorlinks=true,urlcolor=blue
,anchorcolor=blue,citecolor=blue,filecolor=blue,linkcolor=red,menucolor=blue
]{hyperref}

\usepackage{mathtools}
\DeclareMathOperator{\eV}{eV}
\DeclareMathOperator{\ev}{eV} \DeclareMathOperator{\kev}{keV} \DeclareMathOperator{\mev}{MeV}
\DeclareMathOperator{\MeV}{MeV}         \DeclareMathOperator{\km}{km}  \DeclareMathOperator{\s}{s}   \DeclareMathOperator{\erg}{erg}  \DeclareMathOperator{\few}{few} 
      \newcommand{\cL}{{\cal L}} \newcommand{\cM}{{\cal M}}  \newcommand{\cO}{{\cal O}}

\newcommand{\pL}{\left(} \newcommand{\pR}{\right)}      
\newcommand{\beq}{\begin{equation}} \newcommand{\eeq}{\end{equation}}
\newcommand{\bea}{\begin{eqnarray}} \newcommand{\eea}{\end{eqnarray}}

\newcommand{\tenx}[1]{\times 10^{#1}}

\def\lsim{\mathrel{\raise.3ex\hbox{$<$\kern-.75em\lower1ex\hbox{$\sim$}}}}
\def\gsim{\mathrel{\raise.3ex\hbox{$>$\kern-.75em\lower1ex\hbox{$\sim$}}}}

\newcommand{\Eq}[1]{Eq.~(\ref{#1})} \newcommand{\Eqs}[2]{Eqs.~(\ref{#1}) and (\ref{#2})} 
\newcommand{\Sec}[1]{Sec.~\ref{#1}}

\newcommand{\App}[1]{App.~\ref{#1}}

\newcommand{\be}{\begin{eqnarray}}
\newcommand{\ee}{\end{eqnarray}}

\newcommand{\benum}{\begin{enumerate}}
\newcommand{\eenum}{\end{enumerate}}
\newcommand{\bi}{\begin{itemize}}
\newcommand{\ei}{\end{itemize}}

\newcommand{\geff}{G_{\rm eff}}
\newcommand{\GF}{G_{\rm F}}

\newcommand{\neff}{{N_{\rm eff}}}
\newcommand{\dneff}{{\Delta N_{\rm eff}}}
\def\lcdm{$\Lambda$CDM\xspace}

\begin{document}

\preprint{FERMILAB-PUB-19-175-A-T}

\title{Constraining the Self-Interacting Neutrino Interpretation of the Hubble Tension}

\author{Nikita Blinov}
\thanks{ORCID: \url{http://orcid.org/0000-0002-2845-961X}}

\author{Kevin J.~Kelly}
\thanks{ORCID: \url{http://orcid.org/0000-0002-4892-2093}}

\author{Gordan Krnjaic}
\thanks{ORCID: \url{http://orcid.org/0000-0001-7420-9577}}

\author{Samuel D.~McDermott}
\thanks{ORCID: \url{http://orcid.org/0000-0001-5513-1938}}

\affiliation{Fermi National Accelerator Laboratory,  Batavia, IL, USA}

\date{\today}

\begin{abstract}
Large, non-standard neutrino self-interactions have been shown to resolve the $\sim 4\sigma$ tension in Hubble constant measurements 
and a milder tension in the amplitude of matter fluctuations.
We demonstrate that interactions of the necessary size imply the 
existence of a force-carrier with a large neutrino coupling ($> 10^{-4}$) and mass in the keV -- 100 MeV range. This mediator is subject 
to stringent cosmological and laboratory bounds, and we find that nearly 
all realizations of such a particle are excluded by existing data unless it carries spin 0 and couples 
almost exclusively to $\tau$-flavored neutrinos. 
Furthermore, we find that the light neutrinos must be Majorana, and that a UV-complete model requires a non-minimal mechanism to simultaneously generate neutrino masses and appreciable self-interactions.

\end{abstract}

\maketitle

\section{Introduction}
The discrepancy between low-redshift and Cosmic Microwave Background (CMB) determinations of the present-day Hubble parameter, $H_0$, has grown in significance to $\sim 4\sigma$ over several years~\cite{Riess:2016jrr,Shanks:2018rka,Riess:2018kzi,Aghanim:2018eyx,Riess:2019cxk}.
The standard cosmological model, $\Lambda$CDM, may need to be augmented if this ``$H_0$ tension'' is not resolved by observational systematics.
This tension cannot be addressed by modifying $\Lambda$CDM at low redshift~\cite{Vonlanthen:2010cd, Verde:2016wmz, Evslin:2017qdn, Aylor:2018drw};
 adding new physics before recombination seems more promising~\cite{Lesgourgues:2015wza,DiValentino:2017oaw, Poulin:2018dzj, DEramo:2018vss, Poulin:2018cxd, Pandey:2019plg,Escudero:2019gzq,Agrawal:2019lmo}. The solutions in Refs.~\cite{Lesgourgues:2015wza,DiValentino:2017oaw, Poulin:2018dzj, DEramo:2018vss, Poulin:2018cxd, Pandey:2019plg,Escudero:2019gzq,Agrawal:2019lmo} operate at temperatures $\gtrsim 1\;\eV$ to modify the sound horizon and the inferred value of $H_0$.
Low-redshift measurements of the matter density fluctuation amplitude on 8 Mpc scales, $\sigma_8$, also appear to be lower than predicted by $\Lambda$CDM from 
the CMB. This milder ``$\sigma_8$ tension'' is not ameliorated in~\cite{DiValentino:2017oaw, Poulin:2018dzj, DEramo:2018vss,Poulin:2018cxd, Pandey:2019plg, Escudero:2019gzq}.

One resolution to both issues is non-standard neutrino self-interactions~\cite{Cyr-Racine:2013jua,Archidiacono:2013dua,Lancaster:2017ksf,Oldengott:2017fhy,Kreisch:2019yzn}
\be
\label{genericLeff}
{\cal L}_{\rm eff} = \geff (\bar \nu \nu )(\bar \nu \nu),
\ee
where $\geff$ is a dimensionful coupling with flavor indices suppressed. 
If $\geff$ is much larger than the Standard Model (SM) Fermi constant, $\GF$, neutrinos remain tightly coupled to each other until relatively late times. 
  This inhibits their free-streaming, resulting in enhanced power on small scales and a 
  shift in the acoustic peaks of the CMB spectrum relative to \lcdm~\cite{Bashinsky:2003tk}.

The effect of self-interactions is degenerate with other parameters in the CMB fit, including the angular scale of the sound horizon, 
the spectral index and amplitude of primordial fluctuations, and extra radiation.
These degeneracies enable a \emph{preference} for $\geff \gg \GF$ in cosmological data~\cite{Cyr-Racine:2013jua,Lancaster:2017ksf,Oldengott:2017fhy,Kreisch:2019yzn} while relaxing the $H_0$ tension~\cite{Oldengott:2017fhy,Lancaster:2017ksf,Kreisch:2019yzn}.
Ref.~\cite{Kreisch:2019yzn} extended previous analyses, allowing for finite neutrino masses and extra radiation at CMB times. They found that $\geff$ in the ``strongly interacting" (SI$\nu$) or  ``moderately interacting'' (MI$\nu$) regimes
\be
\label{H0-favored}
\geff  \! = 
\begin{cases} 
    & \hspace{-0.2cm} (4.7^{+0.4}_{-0.6} \mev)^{-2} ~~~{\rm(SI}\nu) \\[4pt]
 &\hspace{-0.2cm} (89^{+171}_{-61} \mev)^{-2} ~~~{\rm(MI}\nu)
 \end{cases}
\ee
could simultaneously reduce the $H_0$ and $\sigma_8$ tensions.\footnote{These regions correspond to the \texttt{Planck TT+lens+BAO+$H_0$} datasets. Other dataset combinations considered in Ref.~\cite{Kreisch:2019yzn} prefer similar values 
of $\geff$.} 
Interestingly, the SI$\nu$ cosmology prefers a value of $H_0$ compatible with local measurements at the 1$\sigma$ level, 
even before including local data in the fit.

The range of $\geff$ in \Eq{H0-favored} vastly exceeds the strength of weak interactions, whose coupling is $G_{\rm F} \simeq  (2.9\tenx{5}\mev)^{-2}$. 
We show that this interaction can only arise from the virtual exchange of a
force carrier (``mediator") with  $\cO(\mev)$ mass and appreciable couplings to neutrinos. 
For this mass scale, the effective interaction in \Eq{genericLeff} is valid at energies of order $ \lesssim 100\;\eV$, which prevail during the CMB era.
 However, at higher energies, this mediator is easy to produce on shell, and is subject to stringent cosmological and laboratory bounds.
 
We find that if strong neutrino self-interactions resolve the $H_0$ tension, then:
 \begin{itemize}
   \item {\bf Flavor-universal $\geff$ excluded:}  If $\geff$ is neutrino flavor-universal, 
     both SI$\nu$ and MI$\nu$ regimes in \Eq{H0-favored} are excluded by laboratory searches for rare $K$ decays and neutrinoless double-beta decay.
 \item {\bf MI$\nu$ interactions with $\nu_\tau$ favored:}  Couplings to $\nu_e$, $\nu_\mu$ with
$\geff$ in the range of \Eq{H0-favored} are also excluded, except for a small island
 for $\nu_\mu$ coupling. The only viable scenario involves neutrinos interacting through their $\nu_\tau$ components in the MI$\nu$ regime.
 \item {\bf Vector forces excluded:} Constraints from Big Bang Nucleosynthesis (BBN) exclude most self-consistent vector mediators.
 \item{\bf Dirac neutrinos disfavored:} Mediator-neutrino interactions thermalize the right-handed components of Dirac neutrinos, significantly increasing the number of neutrino species at BBN. This excludes nearly all scenarios except the MI$\nu$ regime with couplings to $\nu_\tau$.
\item {\bf Minimal seesaw models disfavored:} Achieving the necessary interaction strength from a gauge-invariant, UV-complete model, while simultaneously accounting for neutrino masses is challenging in minimal seesaw models.
 \end{itemize}

This work is organized as follows:  \Sec{light-mediator} demonstrates that a light new particle is required to generate the interaction in \Eq{genericLeff} with appropriate strength; \Sec{CosmoBounds} presents cosmological bounds on this scenario;  \Sec{bounds-lab} discusses corresponding laboratory constraints; \Sec{actual-UV}
 shows how \Eq{genericLeff} can arise in UV complete models; finally,  \Sec{conclusion} offers some concluding remarks.

\section{The Necessity of a Light Mediator} \label{light-mediator}

Refs.~\cite{Oldengott:2017fhy,Lancaster:2017ksf,Kreisch:2019yzn, Barenboim:2019tux} assume that all left-handed (LH)
neutrinos undergo $2 \to 2$ flavor-universal scattering described by the interaction in \Eq{genericLeff}. 
The largest detected CMB multipoles correspond to modes that entered the horizon 
when the neutrino temperature was $< 100\;\eV$. This sets the characteristic energy scale of scattering reactions during this epoch: it is important that the form of the Lagrangian in \Eq{genericLeff} is valid at this temperature.
At higher energies, however, this description breaks down. As previously noted in~\cite{Oldengott:2017fhy,Cyr-Racine:2013jua,Lancaster:2017ksf,Kreisch:2019yzn}, the operator in \Eq{genericLeff} is non-renormalizable, and thus is necessarily replaced by a different interaction with new degree(s) of freedom at a scale higher than the $\sim \cO(100\ev)$ energies probed by the CMB (see Ref.~\cite{Cohen:2019wxr} for a review).

The interaction in \Eq{genericLeff} can be mediated by a particle $\phi$ with mass $m_\phi$ and coupling to neutrinos $g_\phi$: 
\beq \label{Lphen}
\cL \supset -\frac12 m_\phi^2 \phi^2 + \frac12( g_\phi^{\alpha \beta} \nu_\alpha \nu_\beta \phi + {\rm h.c.}),
\eeq
where $\nu_\alpha$ are two-component left-handed neutrinos, and we allow for generic flavor 
structure $g_\phi^{\alpha\beta}$ of the interaction. 
In \Eq{Lphen} we have assumed that $\phi$ is a real scalar; our conclusions are unchanged if $\phi$ is CP-odd or complex. 
Vector forces face stronger constraints than scalars, as discussed below.

Using \Eq{Lphen}, we see that the $\nu \nu \to \nu \nu$ scattering amplitude is ${\cal M} \propto g_\phi^2/(m_\phi^2 - q^2)$.
If the momentum transfer $q$ satisfies $|q^2| \ll m_\phi^2$, then $\cM \propto \geff \pL 1 + q^2/m_\phi^2 + \cdots \pR,$ where 
 \be
\label{geff-scalar}
\geff \equiv \frac{g_\phi^2}{m_\phi^2}  = (10~\MeV)^{-2} \left(  \frac{g_\phi}{10^{-1}} \right)^2 
\left(  \frac{\rm MeV}{m_\phi} \right)^2~.
\ee
If $m_\phi^2 \ll |q^2|$, $\cM \propto g_\phi^2/q^2$, leading to qualitatively different energy dependence for neutrino self-interactions;
 this regime was investigated in Refs.~\cite{Forastieri:2015paa, Forastieri:2019cuf}, which found no improvement in the $H_0$ tension.\footnote{Unlike Ref. \cite{Kreisch:2019yzn}, Refs.~\cite{Forastieri:2015paa, Forastieri:2019cuf} fixed $\neff$ and $\sum m_\nu$, but we note that a light mediator would affect multipoles between the first acoustic peak and the diffusion scale. This should be contrasted with the massive mediator case where the self-interaction effects are larger at higher multipoles, allowing for non-standard values of $\neff$ and $\sum m_\nu$ to compensate. A strongly-interacting mode could exist here, but is unlikely to result in a larger value of $H_0$ after accounting for $\neff$ and $\sum m_\nu$ effects, since these impact higher-$\ell$ modes of the CMB spectrum.}
 Thus, we focus on models with a new particle $\phi$ for which $m_\phi^2 \gg |q^2|$ at energy scales relevant to the CMB.

Throughout this epoch, neutrinos are relativistic, so the typical momentum transfer is $|q^2| \sim T_\nu^2$. Eq.~(\ref{geff-scalar}) is valid if $m_\phi \gg T_\nu$. Comparing the values in \Eq{H0-favored} to $\geff$ in \Eq{geff-scalar},
 \be
 \label{mass}
 m_\phi \simeq (4-200)\times | g_\phi | \mev ~.
 \ee
Since perturbativity requires $g_\phi  \lesssim 4\pi$, a new sub-GeV state is required to realize this self-interacting-neutrino solution. 
 Since $T_\nu < 100\ev$ at horizon entry of the highest observed CMB multipoles, 
 the validity of \Eq{genericLeff} in the analyses of~\cite{Cyr-Racine:2013jua,Lancaster:2017ksf,Oldengott:2017fhy,Kreisch:2019yzn, Barenboim:2019tux} requires $m_\phi \gtrsim \kev$ (as noted in~\cite{Kreisch:2019yzn}).
   From \Eq{mass}, this translates to
 \be
 \label{min-coupling}
 m_\phi \gtrsim \kev ~~\implies~~ |g_\phi| \gtrsim 10^{-4}~.
 \ee
 This bounds the allowed ranges of $m_\phi$ and $g_\phi $. Note that \Eq{mass} precludes 
 the new self-interactions from being described within Standard Model effective theory with 
 no light states below the weak scale~\cite{Gavela:2008ra}.
 
Finally, we note that \Eq{Lphen} is not gauge-invariant at energies
above the scale of electroweak symmetry breaking (EWSB). We explore UV completions in Sec.~\ref{actual-UV}.

\begin{figure*}
\includegraphics[width=3.3in,angle=0]{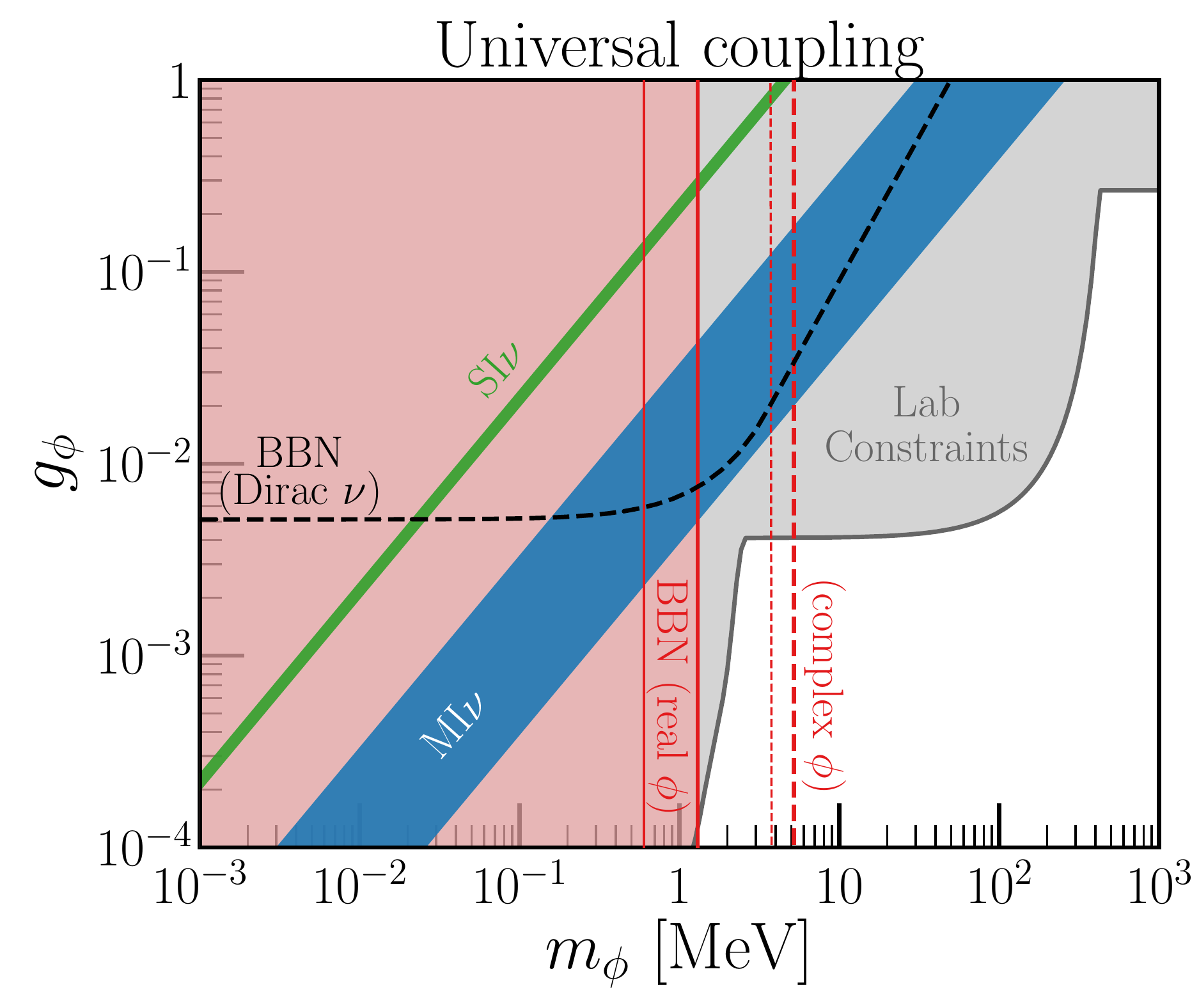}~~
\includegraphics[width=3.3in,angle=0]{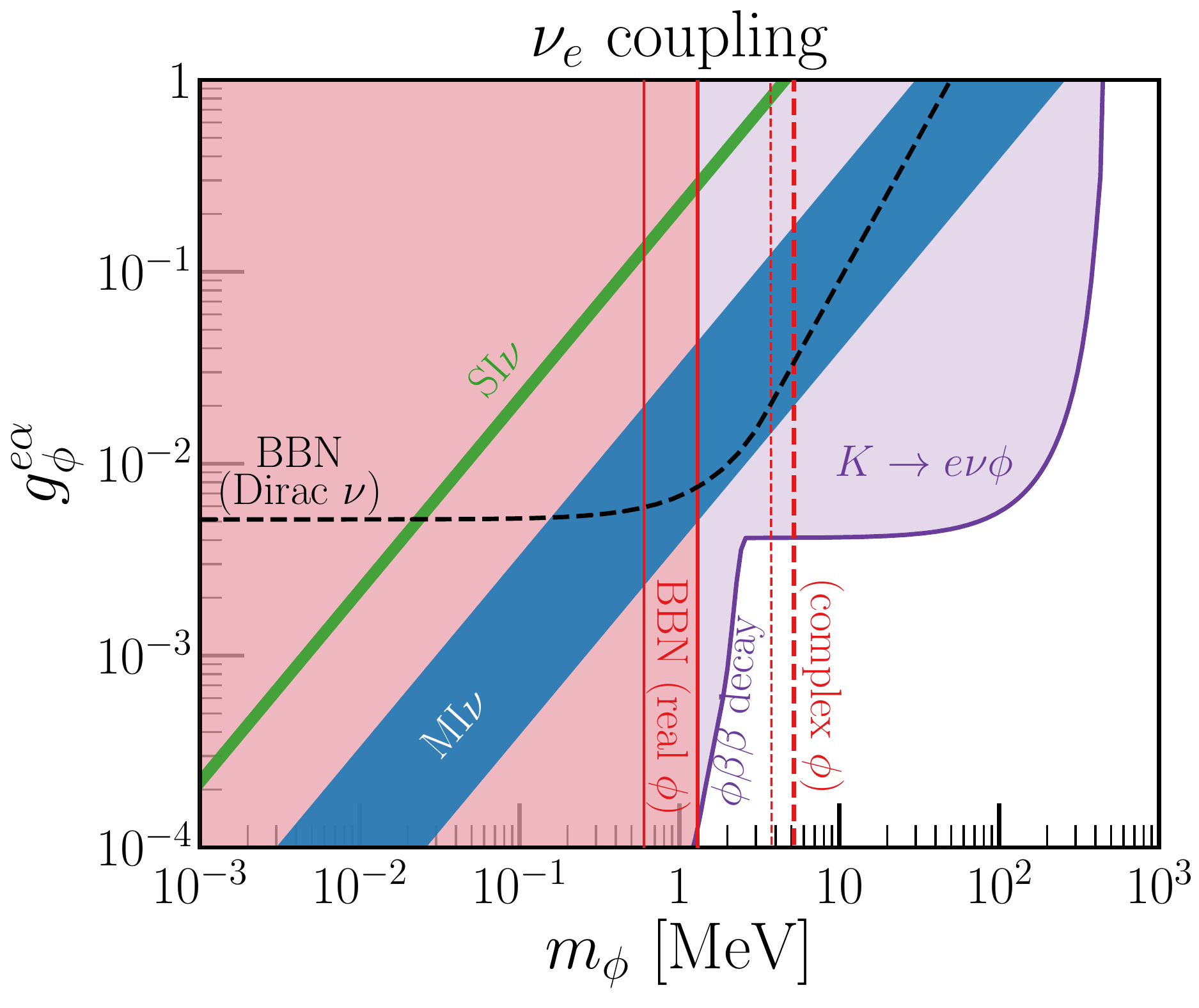}\\
\includegraphics[width=3.3in,angle=0]{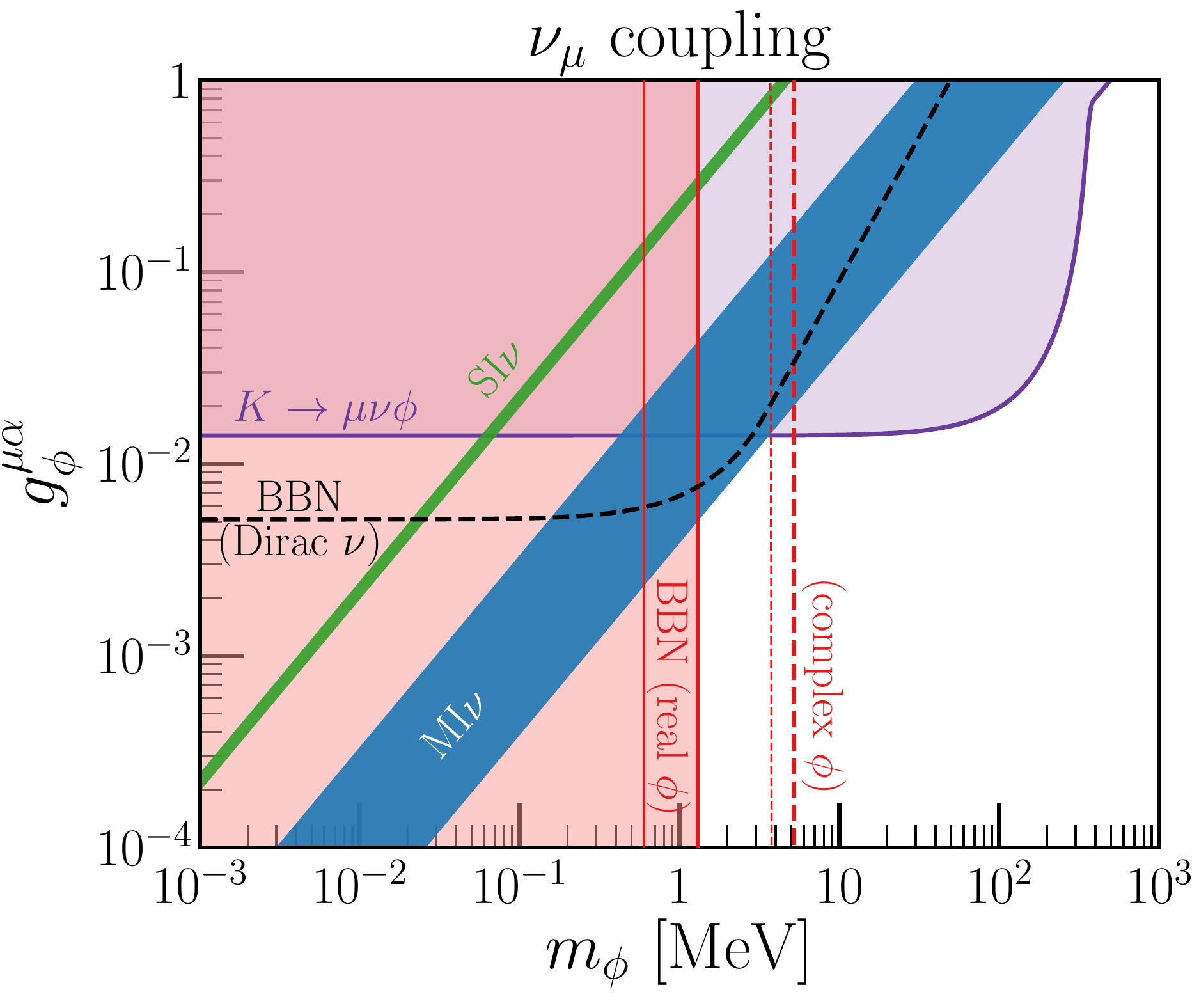}~~
\includegraphics[width=3.3in,angle=0]{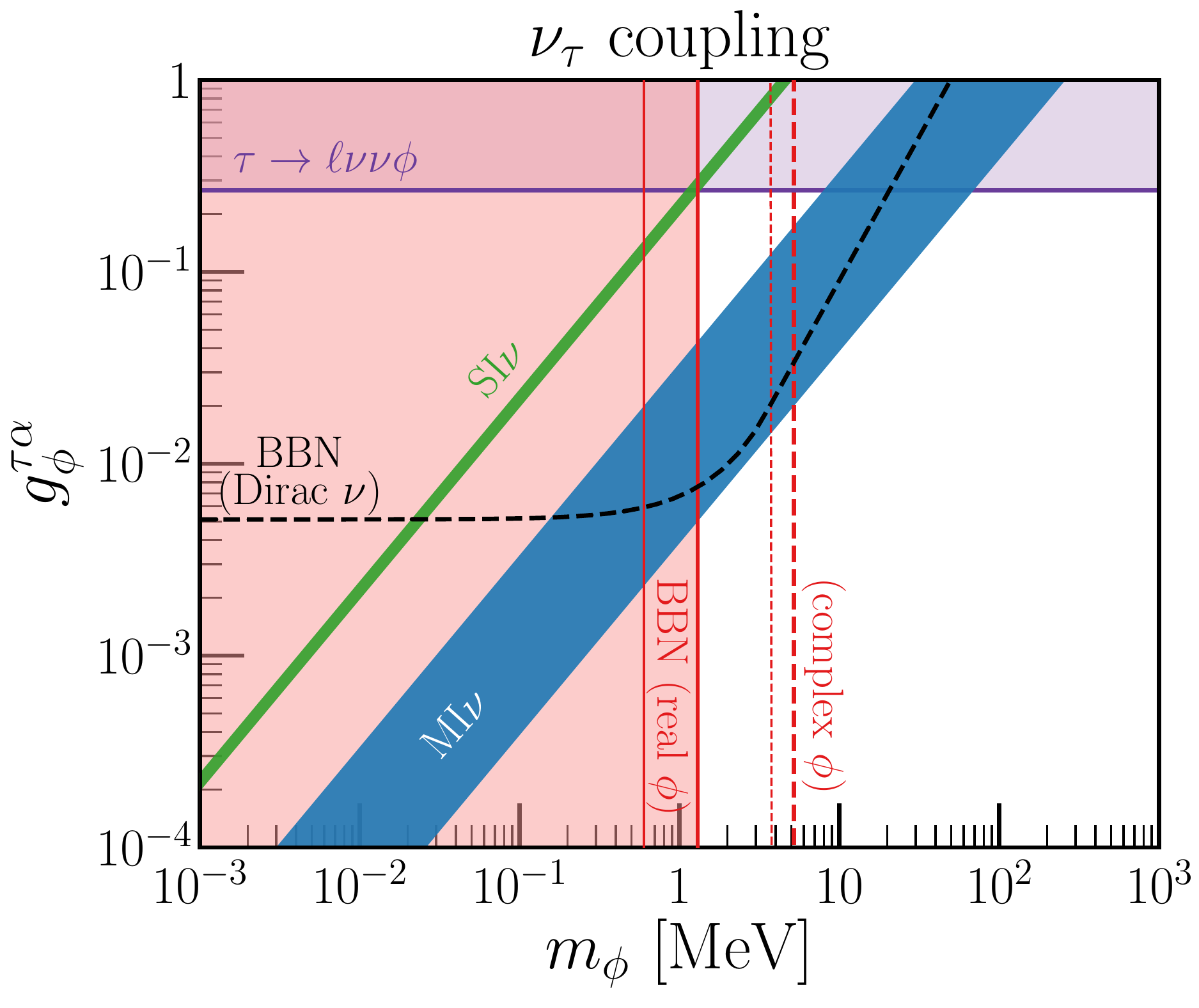}\\
\caption{Bounds (shaded regions) on light neutrino-coupled mediators with flavor-universal couplings (top-left), and 
  flavor-specific couplings to $\nu_{e}$ (top-right),  $\nu_\mu$ (bottom-left), and $\nu_\tau$ (bottom-right).
The bands labeled MI$\nu$ and SI$\nu$ are the preferred regions from
\Eq{H0-favored} \cite{Kreisch:2019yzn} translated into the $g_\phi$-$m_\phi$ plane. Also shown are
constraints from $\tau$ and rare meson decays~\cite{Blum:2014ewa,Berryman:2018ogk,Kelly:2019wow,Krnjaic:2019rsv}, double-beta decay experiments~\cite{Agostini:2015nwa, Blum:2018ljv, Brune:2018sab} (purple), and BBN (red). 
We combine the $\tau$/meson decay and double-beta decay constraints as ``Lab Constraints'' in the upper-left panel.
BBN yields depend on the baryon density $\eta_b$; thick (thin) lines correspond to the SI$\nu$ (MI$\nu$) preferred values of $\eta_b$. 
Nucleosynthesis constraints are stronger for complex scalar mediators (dashed red) than for real scalars (solid red).
If neutrinos are Dirac, their right-handed components equilibrate before BBN above the dashed black line.
 }
\label{constraints}
\end{figure*}

\section{Cosmological Bounds}
\label{CosmoBounds}

Successful predictions of BBN provide a powerful probe of additional light species. New particles in thermal equilibrium with neutrinos increase 
the expansion rate during BBN as extra relativistic degrees of freedom or by heating neutrinos relative to photons. 
Away from mass thresholds, both effects are captured by a constant shift in $\neff$, 
the effective number of neutrinos.
We find that the observed light element abundances constrain $\dneff < 0.5$ ($0.7$) at 95\% CL for the 
SI$\nu$- (MI$\nu$-) preferred values of the baryon density, as detailed in App.~\ref{dneffsec}.

We emphasize that large $\dneff\simeq 1$ at CMB times is crucial for the MI$\nu$ and SI$\nu$ results~\cite{Kreisch:2019yzn}. 
Since BBN does not prefer large $\neff$, the self-interacting neutrino framework requires an injection of 
energy between nucleosynthesis and recombination, e.g., via late equilibration of a dark sector~\cite{Berlin:2017ftj}.
Such scenarios may face additional constraints. To remain model-independent, 
we only consider the implications of BBN for the mediator (and right-handed neutrinos if they are Dirac particles) needed 
to implement strong neutrino self-interactions.

\subsection{Mediators and \texorpdfstring{$\dneff$}{Delta Neff}}
\label{mediatorneff}

\Eq{Lphen} induces $\phi \leftrightarrow \nu\nu$ decays and inverse decays, 
which can equilibrate $\phi$ with neutrinos before neutrino-photon decoupling at $T_{\rm dec} \sim 1-2 \;\MeV$. 
Here we show that this necessarily happens for mediators that realize $\geff$ in \Eq{H0-favored}.
Annihilation and scattering processes also 
contribute, but the corresponding rates are suppressed by additional powers of  $g_\phi$.

~\\ \noindent {\bf Vector Mediators:} If \Eq{genericLeff} arises from a vector particle $\phi_\mu$ with mass $m_\phi$, then at energies above $m_\phi$ ${\cal L} \longrightarrow \frac12 m_\phi^2 \phi^\mu \phi_\mu+ \pL g_\phi \phi_\mu \nu^\dagger \bar \sigma^\mu \nu + {\rm h.c.} \pR,$ where $g_\phi$ is the gauge coupling. 
$\phi_\mu$ equilibrates before $T_{\rm dec}$ via $\nu \nu \leftrightarrow \phi$ 
if the corresponding thermally-averaged rate $\Gamma_{\nu\nu\to \phi}$ exceeds Hubble 
when $T=\max(T_{\rm dec}, m_\phi)$:
\be
\frac{\Gamma_{\nu \nu \to \phi}}{H}  \sim \frac{g_\phi^2 m_\phi^2 M_{\rm Pl} }{ \max(T_{\rm dec},m_\phi)^3} > 10^8 \,  \frac{\geff}{(10\, \MeV)^{-2}},
\ee
where $M_{\rm Pl} = 1.22 \times 10^{19}$ GeV and we have used \Eqs{geff-scalar}{min-coupling}.   
This reaction is in equilibrium for all values of couplings and masses of interest. 
As a result, $\phi_\mu$ has a thermal number density at $T_{\rm dec}$ in both MI$\nu$ and SI$\nu$ scenarios. 
Counting degrees of freedom, we find $\dneff = (8/7)(3/2) \simeq 1.7$ assuming $\phi_\mu$ remains
relativistic throughout BBN; if $\phi_\mu$ becomes non-relativistic between $T_{\rm dec}$ and 
the end of BBN, then $\dneff \approx 2.5$. Thus, $\phi_\mu$ must become non-relativistic well before $T_{\rm dec}$.
Ref.~\cite{Escudero:2019gzq} found that Boltzmann suppression for massive vectors
is effective for $m_\phi > 10 \, \mev$ (95\% CL). Using \Eq{geff-scalar},  
this requires $g_\phi \gtrsim \cO(0.1)$, which is excluded 
in all theoretically-consistent (or anomaly-free) vector models with neutrino couplings~\cite{Bauer:2018onh,Farzan:2015hkd}. 
Anomaly-free vectors, such as those coupled to lepton-family-number currents,
would introduce large $\bar{\nu}\nu \bar{e}e$ interactions which would likely spoil the CMB fit.
 
~\\ \noindent {\bf Scalar Mediators:} Similarly, any scalar mediator $\phi$ that realizes $\geff$ from \Eq{Lphen} with $g_\phi \gtrsim 10^{-4}$ (required by \Eq{min-coupling}) also has a thermal abundance at $T_{\rm dec}$. Relativistic scalars 
in equilibrium with neutrinos contribute $\dneff = 0.57\;(1.1)$ for a real (complex) $\phi$, which has one (two) degree(s) of freedom. 
The $\phi$ density must become Boltzmann-suppressed before neutrino-photon decoupling, leading 
to a lower limit on $m_\phi$. We use \texttt{AlterBBN} 2.1~\cite{Arbey:2011nf,Arbey:2018zfh} as described in App.~\ref{dneffsec} to obtain lower bounds (95\% CL)
\be
  \label{scalarBBNbound}
  m_\phi > 
  \begin{cases}
1.3 \, \mev ~~ (\text{real scalar})\\
5.2 \, \mev ~~ (\text{complex scalar})
\end{cases},
  \ee 
  for the SI$\nu$ preferred values of the baryon density (corresponding MI$\nu$ bounds are somewhat weaker -- see App.~\ref{dneffsec}).
SI$\nu$ and MI$\nu$ BBN bounds are presented in Fig.~\ref{constraints} as thick and thin red vertical lines, respectively.

\subsection{Constraining Dirac Neutrinos}
\label{dirac}

If neutrinos are Dirac all neutrino masses arise from the interaction ${\cal L}_{\rm Dirac} \supset y_\nu H L \nu_R \to m_\nu \nu \nu_R$, where $m_\nu \equiv y_\nu v/\sqrt{2},$ $H$ is the Higgs doublet, $L = (\nu,\; \ell)^T$ is a lepton doublet,
$\nu_R$ is a right-handed neutrino (RHN), and flavor indices have been suppressed. The Weyl fermions
 $\nu$ and $\nu_R$ become Dirac partners after EWSB and acquire identical masses.
In the SM alone, the Yukawa coupling $y_\nu \sim 10^{-12} (m_\nu/ {\rm 0.1 \, eV})$ is insufficient to thermalize right-handed states, so relic neutrinos consist of left-handed neutrinos and right-handed antineutrinos~\cite{Dolgov:1994vu}.
 
The interactions in \Eq{H0-favored} are much stronger than
 the weak force at late times, so $\phi$ and $\nu_R$ can both thermalize. Approximating 
 the RHN production rate as  $\Gamma_{\phi \to \nu \nu_R} \simeq (m_\nu/m_\phi)^2 \Gamma_{\phi \to \nu   \nu }$,
for $m_\nu  = 0.1$ eV we have
\be 
\frac{  \Gamma_{\phi \to \nu \nu_R} }{H}  \simeq  \frac{g_\phi^2 m_\nu^2 M_{\rm Pl}}{m_\phi^3} 
= 10^6 \frac{  \geff  }{(10 \mev)^{-2}} \frac{\rm MeV}{m_\phi} ,~~
\ee
where $T = m_\phi \gtrsim T_{\rm dec}$ is the temperature at which RHN production is maximized relative to $H$. See App.~\ref{dirnusec} for more details.

Neutrino oscillation results require that at least two of the light neutrinos are massive, with one heavier than $\sim 10^{-2}$ eV and one heavier than $\sim 10^{-1}$ eV~\cite{Esteban:2018azc}. For all values of $m_\phi$ we consider in the SI$\nu$ range, at least one RHN will thermalize before BBN, leading to $\Delta \neff \gtrsim 1$. We therefore assume that neutrinos are Majorana particles for the remainder of this work.

\subsection{Secret Neutrino Interactions}
The $\neff$ bounds considered here can, in principle, be evaded by ``secret" interactions 
which are communicated to active neutrinos via mixing with a light sterile neutrino, which couples directly
to a mediator. In these scenarios the active-sterile mixing angle is suppressed at early times by plasma effects, 
 but can become large at later times when the universe is cooler~\cite{Abazajian:2001nj,Hannestad:2013ana,Hannestad:2012ky,Saviano:2013ktj}. 
 The mixing angle may be smaller than
  $\sim 10^{-9}$ for $T \gtrsim 50$ keV when BBN ends (to avoid thermalization)
and subsequently grow to $\sim {\cal O}(1)$ by $T\sim 100$ eV (to enable a large active neutrino self-interaction during the CMB era, thereby resolving the $H_0$ tension). 
This sharp transition over a narrow temperature range requires significant fine tuning of the 
active-sterile mass-splitting and a large lepton asymmetry. See App.~\ref{SecretNu} for a discussion.

\section{Laboratory Bounds } \label{bounds-lab}

Because terrestrial experiments routinely reach energies above the MeV scale, the model of \Eq{Lphen} is well constrained.
We focus on scalar mediators, commenting on pseudoscalars in \Sec{actual-UV}. Laboratory constraints arise from: 
    \medskip

\noindent {\bf Double Beta Decay:} If $g_\phi^{ee} \neq 0$ and $\phi$ is lighter than the $Q$-value of a double-beta-decaying nucleus, the process $(Z, A) \to (Z+2, A)\, e^- e^- \phi$ may occur,
  contributing to measured $2\nu\beta\beta$
  rates. Measurements constrain $|g_\phi^{ee}| \lesssim 10^{-4}$ if $m_\phi \lesssim 2$ MeV~\cite{Agostini:2015nwa, Blum:2018ljv, Brune:2018sab}, shown in the top row of Fig.~\ref{constraints}. 

    \medskip
    
\noindent {\bf Meson Decays:} Nonzero $g_\phi^{\alpha\beta}$ can allow for meson decays $\mathfrak{m}^\pm \to \ell_\alpha^\pm \nu_\beta \phi$ if $m_\phi < m_\mathfrak{m} - m_{\ell_\alpha}$~\cite{Blum:2014ewa,Ng:2014pca,Ioka:2014kca,Berryman:2018ogk,Kelly:2019wow,Krnjaic:2019rsv}. $\mathrm{Br}(K^+ \to e^+ \nu_e)/\mathrm{Br}(K^+\to \mu^+ \nu_\mu) = \left(2.416 \pm 0.043\right) \times 10^{-5}$ constrains $g_{\phi}^{e\beta}$ as shown in the top row of Fig.~\ref{constraints}~\cite{Lessa:2007up,Fiorini:2007zzc}.
$\mathrm{Br}(K^+ \to \mu^+ \nu_\mu \nu \bar{\nu}) < 2.4 \times 10^{-6}$~\cite{Tanabashi:2018oca} constrains $g_\phi^{\mu\alpha}$, shown by the purple region in bottom-left panel of Fig.~\ref{constraints}. 
  
    \medskip
    \noindent {\bf  $\tau$ Decays:} The decay $\tau^- \to \ell_\beta \overline{\nu}_\beta \overline{\nu}_\tau \phi$ constrains $g_\phi^{\tau\tau}$. Ref.~\cite{Lessa:2007up} found $g_\phi^{\tau\tau} \lesssim 0.3$ for light $\phi$, depicted as a purple band in the bottom-right panel of Fig.~\ref{constraints}.

~\\

Fig.~\ref{constraints} summarizes our findings: values of $\geff$ from \Eq{H0-favored} 
favored by the $H_0$ tension are excluded if 
$\phi$ couples universally to all neutrinos (top-left),
which was explicitly considered in Refs.~\cite{Cyr-Racine:2013jua,Lancaster:2017ksf,Oldengott:2017fhy,Kreisch:2019yzn,Barenboim:2019tux}, 
or (in the SI$\nu$ solution) if $\phi$ couples predominantly to 
 $\nu_e$ or $\nu_\mu$ (top-right/bottom-left panels, respectively). Similarly, we can exclude the possibility that $\phi$ couples to any single mass-eigenstate neutrino,
 since the $\nu_e$- and $\nu_\mu$-composition of each mass eigenstate is similar. Moreover, in this case, the collisional Boltzmann equations would be much more complicated to solve (different eigenstates will start to free-stream at different times), and the results of Refs.~\cite{Cyr-Racine:2013jua,Lancaster:2017ksf,Oldengott:2017fhy,Kreisch:2019yzn, Barenboim:2019tux} may not apply.

 However, a {\it flavor-restricted} coupling leads to approximately the same neutrino mass-eigenstate interactions as in Refs.~\cite{Cyr-Racine:2013jua,Oldengott:2017fhy,Lancaster:2017ksf,Kreisch:2019yzn, Barenboim:2019tux}, since the flavor eigenstates are well-mixed in the mass basis. A $\nu_\tau$-only coupling, in which the matrix $g_{\alpha \beta}$ is zero except for $g_{\tau\tau}$, is potentially viable since $\tau$ decays are less constraining than meson decays. 
 Thus, we are unable to fully exclude an interaction $G_{\rm eff}^\tau \bar \nu_\tau \nu_\tau \bar \nu_\tau \nu_\tau$.
 
  In this case, $G_{\rm eff}^\tau =A \times G_{\rm eff}$ for $\geff$ defined in \Eq{geff-scalar} and $A \sim \cO(1)$ is a constant that accounts for the reduced scattering probability of each mass eigenstate. 
Because mixed mass-eigenstate vertices are possible in this scenario, there are additional diagrams compared to the mass-diagonal case. 
For this reason, we caution that the effect on the CMB anisotropies of flavor-specific neutrino self-interactions can be mildly different than that considered in Refs.~\cite{Cyr-Racine:2013jua,Lancaster:2017ksf,Oldengott:2017fhy,Kreisch:2019yzn, Barenboim:2019tux}.
Nonetheless, we expect that the preferred coupling range should shift slightly {\it up} relative to the flavor-universal case; a complete study is necessary to know how this affects the full SI$\nu$ range.
The MI$\nu$ range is still allowed in a $\tau$-flavor-only scenario, though a dedicated study is needed.

\section{Ultraviolet Completions} \label{actual-UV}

In this section we 
consider models of Majorana neutrinos with an additional particle $\phi$, specifically the type-I and II seesaw mechanisms. 
In both, we find the resulting $\phi \nu \nu$ coupling is suppressed by factors of the light neutrino mass. 
In these minimal models, it is therefore impossible to 
simultaneously generate neutrino masses and a large enough $\geff$ to address the $H_0$ tension. 

We note that the coupling of $\phi$ to LH neutrinos in Eq.~\ref{Lphen} 
violates lepton number in analogy to neutrino masses, so it is a
compelling possibility to relate these phenomena. The SM Lagrangian preserves lepton number, so the 
the scale $f$ of lepton-number violation must arise 
from new interactions. In type-I models, $f$ is related to the 
RH neutrino mass, while in type-II it is proportional to the Higgs-triplet mixing parameter~\cite{Cai:2017jrq}. 
The interaction of $\phi$ with the neutrino sector occurs through the combination 
$f + \lambda \phi$, where $\lambda$ is a coupling constant.
While the relation of $\lambda$ to the neutrino-masses is model-dependent, 
the interaction with $\phi$ takes on the universal form $  g_\phi \approx \lambda m_\nu/f \Rightarrow \geff \sim \lambda^2 m_\nu^2/(m_\phi^2 f^2)$ in both type-I and type-II seesaw scenarios.
Realizing
 $\geff \approx \left(4 - 300\ \mathrm{MeV}\right)^{-2}$ 
requires $f\sim 10^3 m_\nu \sim 10 \, \eV$. 
In the type-I model, this scale sets the mass of the RHNs, which thermalize before BBN and contribute to $\dneff$ as in the 
Dirac case discussed in Sec.~\ref{dirac}. 
In type-II models this scale is bounded by non-observation of rare lepton-number-violating processes~\cite{Akeroyd:2009nu,Dinh:2012bp,Dev:2017ouk,Cai:2017jrq}.
Therefore, minimal scenarios where the same seesaw generates neutrino masses and the operator in \Eq{genericLeff} with 
the magnitude in \Eq{H0-favored} are not possible. 

These arguments also apply to the Majoron, the Nambu-Goldstone boson of lepton-number breaking~\cite{Chikashige:1980ui,Chikashige:1980qk,Schechter:1981cv}. In these models, $\phi$ is a pseudoscalar particle, but its coupling to neutrinos is still suppressed by $m_\nu/f$. 
However, the bounds we considered still apply, because all limits derive from relativistic neutrinos, for which there is no distinction between scalar and pseudoscalar.

Finally, we note that large $\geff$ can be obtained using \emph{separate} seesaw mechanisms to generate the neutrino masses and the $\phi \nu\nu$ 
interaction -- we can use the type-I Seesaw for the light neutrino masses and the type-II seesaw mechanism can produce large $g_\phi^{\tau\tau}$ (as long as it does not contribute to $m_\nu$). The size of $g^{\tau\tau}_\phi$ decouples from the neutrino masses.

\section{Concluding Remarks}
\label{conclusion}

We have shown that the self-interacting neutrino explanation of the $H_0$ tension requires the existence of a light $\sim$ MeV-scale mediator, subject to stringent cosmological and laboratory bounds. Consequently,  
for
both the SI$\nu$ and MI$\nu$ regimes in \Eq{H0-favored},
the flavor-universal interactions considered in Refs. \cite{Cyr-Racine:2013jua,Lancaster:2017ksf,Oldengott:2017fhy,Kreisch:2019yzn} are robustly excluded by BBN-only bounds on $\Delta N_{\rm eff}$ and by laboratory searches
for rare $K$ decays and neutrinoless double-beta decay; the SI$\nu$ regime is excluded for all flavor structures. 

Intriguingly, we find that flavor-dependent variations of the MI$\nu$ regime may viably resolve the $H_0$ tension if 
a $\sim$ 10 MeV scalar mediator with large coupling interacts almost exclusively with $\nu_\tau$ or 
$\nu_\mu$ (though there is little parameter space for $\nu_\mu$ coupling). 
A dedicated exploration of the $\tau$-only scenario  
is necessary to determine if the preferred region to resolve the $H_0$ tension persists without running afoul of laboratory measurements. Our results also motivate exploration of the ``intermediate'' mediator-mass regime, where neutrino scattering is relevant for a partial range of redshifts explored by the CMB. 

However, realizing such strong, flavor specific interactions in 
UV-complete, gauge-invariant models is 
 challenging. We find that sufficiently strong interactions cannot arise in models that generate 
neutrino masses via a single type-I or -II seesaw mechanism: the resulting neutrino-scalar coupling is suppressed by factors of $m_\nu/f$ 
where $f \gg m_\nu$ is the appropriate seesaw scale. A compelling and viable model remains to be found.

\begin{acknowledgments}  
  We thank Andr\' e   de Gouv\^ ea, Francis-Yan Cyr-Racine, Joshua Isaacson, Martina Gerbino, Stefan H\"{o}che, Matheus Hostert, Joachim Kopp, Massimiliano Lattanzi, Shirley Li, Kohta Murase, Jessica Turner, and Yue Zhang for helpful conversations. This manuscript has been authored by Fermi Research Alliance, LLC under Contract No. DE-AC02-07CH11359 with the U.S. Department of Energy, Office of Science, Office of High Energy Physics.
\end{acknowledgments}

\appendix

\section{Calculating \texorpdfstring{$\Delta N_{\rm eff}$}{Delta Neff}}
\label{dneffsec}

The Hubble expansion rate at the time of BBN is sensitive to
the energy density in neutrinos and other relativistic species when photon temperatures are below an MeV~\cite{Serpico:2004nm,Boehm:2013jpa,Nollett:2014lwa}.
New relativistic particles or an injection of energy into the Standard Model neutrino bath (via the decays or annihilations of a $\nu$-coupled species) 
can increase the Hubble expansion rate at this time.
A larger expansion rate at the time of BBN modifies the neutron-to-proton ratio and the freeze-out of 
deuterium-burning reactions, leading to larger yields of Helium-4 and Deuterium.
The primordial abundances of these elements are measured in pristine gas clouds to be 
$Y_p = 0.2449\pm 0.004$~\cite{Aver:2015iza} and $10^5$ D/H $= 2.527 \pm 0.03$~\cite{Cooke:2017cwo}, respectively. 
These observations are in good agreement with predictions assuming only Standard Model particle content at the time of BBN~\cite{Cyburt:2015mya}.

We use the measurements of $Y_p$ and D/H 
to constrain modifications to the expansion rate using the BBN Boltzmann code \texttt{AlterBBN} 2.1~\cite{Arbey:2011nf,Arbey:2018zfh}. 
We follow the Monte Carlo procedure outlined in Ref.~\cite{Berlin:2019pbq} to estimate theoretical uncertainties from nuclear reaction rates.
In deriving limits we marginalize over a Gaussian prior on the baryon density $\eta_b$ corresponding to the best-fit points of $\Lambda$CDM+$\neff$+$\sum m_\nu$~\cite{Aghanim:2018eyx}, 
SI$\nu$ and MI$\nu$~\cite{Kreisch:2019yzn}.
Using the results of Refs.~\cite{Steigman:2006nf,Cyburt:2015mya} to 
convert $\Omega_b h^2$ to $\eta_b$, we find that $10^{10}\eta_b =6.086\pm 0.066$, $6.146 \pm 0.082$, and $6.248\pm 0.082$ in 
$\Lambda$CDM+$\neff$+$\sum m_\nu$ (\texttt{TT+lowl+lowE+BAO} data combination), SI$\nu$ and MI$\nu$ (\texttt{TT+lens+BAO+H0}), respectively. 
Note that the best fit values of $\eta_b$ in these cosmologies are all compatible with the range $5.8 <10^{10} \eta_b <  6.6$ (95 \% CL) extracted only from BBN data~\cite{Tanabashi:2018oca}. 

If the new particles remain relativistic throughout nucleosynthesis, their modification of the Hubble rate is 
specified by a constant shift of the number of relativistic degrees of freedom, $\neff$. 
We find that the observed values of $Y_p$ and D/H favor the values of 
$\neff$ shown in Tab.~\ref{tab:neff_bounds} for the different cosmologies.
Note that the $\Lambda$CDM result is higher than reported in, e.g., Ref.~\cite{Cyburt:2015mya}, 
and it has a smaller uncertainty. This is because of updated D burning 
rates and observed abundances used in our analysis; for further discussion, see Ref.~\cite{Berlin:2019pbq}. 
These improvements actually weaken the BBN constraint compared to Ref.~\cite{Cyburt:2015mya} 
because of the slight preference for $\neff > 3$, which is driven by the slight underprediction of D/H compared 
to the observed value.
The upper limits 
  in SI$\nu$ and MI$\nu$ are weaker still, due to their larger central values and uncertainties of the baryon density: a larger 
  baryon density reduces the Deuterium yield, which can be compensated by increasing $\neff$~\cite{Cyburt:2015mya}.
Different choices of datasets feature slightly different best-fit values and uncertainties for $\Omega_b$ in 
the SI$\nu$ and MI$\nu$ cosmologies~\cite{Kreisch:2019yzn}. For MI$\nu$, other data sets prefer slightly lower values of $\Omega_b$, 
which would increase the D/H yield and lead to stronger constraints on $\neff$ (and light thermal particles as discussed below). This makes 
our choice of \texttt{TT+lens+BAO+H0} conservative. In contrast, 
the SI$\nu$ fit that uses \texttt{TTTEEE + lens + H0} has a larger $\Omega_b$; this would weaken the constraints on extra relativistic species. However, 
even this fit has a smaller value of $\Omega_b$  than the MI$\nu$ result discussed above (and shown in Tab.~\ref{tab:neff_bounds}); thus we expect the BBN 
  constraints for \texttt{TTTEEE + lens + H0} to lie between the SI$\nu$ and MI$\nu$ results given in Tab.~\ref{tab:neff_bounds}. 
We also note that our $\Lambda$CDM+$\neff$+$\sum m_\nu$ value of $\Omega_b$ does not include the local $H_0$ measurement in the fit, since 
the CMB-only $H_0$ is incompatible with the local measurement. Combining these (inconsistent) data sets gives a 
preferred value of $\Omega_b$ that is close to that of the MI$\nu$ cosmology~\cite{Kreisch:2019yzn}, and would therefore 
result in BBN bounds similar to the MI$\nu$ results.

The upper bounds in Tab.~\ref{tab:neff_bounds} apply to light particles that are fully relativistic 
at the time of nucleosynthesis. If these particles have masses at the MeV scale, 
then their decays or annihilations can heat the neutrinos relative to photons; if they 
are much heavier, they transfer their entropy to neutrinos while those are still in equilibrium with photons.
The resulting change in neutrino temperature and the corresponding $\dneff$ can 
be estimated assuming instantaneous neutrino-photon decoupling at $T_{\rm dec} \approx 1 - 2\;\mev$ 
and by using entropy conservation~\cite{Boehm:2012gr}. This crude estimate along with Tab.~\ref{tab:neff_bounds} suggests that neutrino-coupled scalars 
in the SI$\nu$ cosmology with $m_\phi \lesssim 2-7$ MeV should be incompatible with the observed light element abundances, where the range corresponds to varying $T_{\rm dec}$ and the number of scalar degrees of freedom from 1 (real scalar) to 2 (complex scalar). 
The results of a full calculation using \texttt{AlterBBN} (which does not make these approximations), shown in Tab.~\ref{tab:mphi_bounds}, 
are compatible with this estimate.

\begin{table}[t]
  \centering
  \def\arraystretch{1.3}
\begin{tabular}{|c|c|c|}
  \hline
Model & $\neff$ & 95\% upper limit  \\
\hline
$\Lambda$CDM+$\neff$+$\sum m_\nu$ & $3.19 \pm 0.135$ & $3.47$ \\ 
SI$\nu$ & $3.27\pm 0.14$ & $3.56$ \\
MI$\nu$ & $3.43\pm 0.13$ & $3.72$ \\
\hline
\end{tabular}
\caption{Preferred values and upper limits on the effective number of neutrino species, $\neff$,  
from primordial nucleosynthesis. The cosmological models differ through their values of 
the baryon density parameter $\eta_b$ (and its uncertainty) determined from the CMB power spectrum as described in the text.
\label{tab:neff_bounds}}
\end{table}

\begin{table}[t]
  \centering
  \def\arraystretch{1.3}
\begin{tabular}{|c|c|c|}
  \hline
  \multirow{3}{*}{Model}        & \multicolumn{2}{|c|}{95\% CL lower bound on $m_\phi$ (MeV)}\\
  \cline{2-3}
               &  ~~~~~~~real $\phi$ ~~~~~~~ & complex $\phi$ \\
\hline
$\Lambda$CDM+$\neff$+$\sum m_\nu$ & $2.3$ & $6.3$\\ 
SI$\nu$ & $1.3$ & $5.2$\\
MI$\nu$ & $0.6$ & $3.7$ \\
\hline
\end{tabular}
\caption{Lower limits on mediator mass $m_\phi$ from primordial nucleosynthesis, assuming 
$\phi$ was in thermal equilibrium before BBN. The cosmological models differ through their values of 
the baryon density parameter $\eta_b$ (and its uncertainty) as determined from the CMB power spectrum.
\label{tab:mphi_bounds}}
\end{table}

We note that we do not use the Planck limit on $\dneff$~\cite{Aghanim:2018eyx}, since that result assumes free-streaming neutrinos. 
In fact, a large $\dneff\simeq 1$ at the time of the formation of the CMB is crucial for the successful fits of MI$\nu$ and SI$\nu$ cosmologies to
the observed CMB power spectrum~\cite{Kreisch:2019yzn}. The 
concordance of the CMB and BBN measurements of $\dneff$ within the self-interacting neutrino framework thus seems to require an injection of 
energy between nucleosynthesis and recombination; we remain agnostic on this point.
Given this discussion, we conservatively only apply a constraint on $\dneff$ from considerations of BBN physics alone.
We further point out that the contribution of a scalar to $\dneff$ at the 95\% CL limit from our BBN analysis is also compatible with the extra radiation density at the best fit point in Ref.~\cite{Kreisch:2019yzn} at only the $2\sigma$ level. In order to be in better agreement with $\dneff \simeq 1.02 \pm 0.29$ for the SI$\nu$ mode, new (semi) relativistic degrees of freedom may need to come into equilibrium with the neutrino bath after BBN is complete~\cite{Hannestad:2012ky,Berlin:2017ftj}.

In deriving constraints from light element abundances we have ignored the possible influence of neutrino self-interactions and 
potentially large neutrino mass.
However, these effects have a negligible impact on the yields. 
For example, large neutrino self-interactions at the time of nucleosynthesis maintain thermal equilibrium for the three neutrino species, and 
affect flavor oscillations. 
In the standard cosmology, the non-instantaneous decoupling of weak interactions leads to a small 
shift to $\neff$ and spectral distortions of the neutrino distributions. 
The former effect is already taken into account in public BBN codes by setting $\neff = 3.046$, while the latter 
has a tiny effect on the abundances of $Y_p$ and D/H~\cite{Mangano:2005cc}. 
  Thus, if the standard spectral distortions are driven to zero by the novel self-interactions considered here, then their impact would still be unobservable. 
  Similarly, the shift to $\neff$ from the presence of non-standard interactions is also much smaller than a percent~\cite{deSalas:2016ztq}.
  Another important parameter in the CMB fit is the neutrino mass, which, along with interaction-induced matter potential, 
can affect flavor oscillations during BBN. However, oscillations have a negligible impact on the abundances~\cite{deSalas:2016ztq}. 
We therefore conclude that the constraints from nucleosynthesis derived above are sensitive to $\neff$ and $\eta_b$, 
but not to other aspects of the strongly-interacting neutrino model.

\section{Dirac Neutrino Thermalization}
\label{dirnusec}

If neutrinos are Dirac fermions, their right-handed (RH) components must not come into equilibrium with the Standard Model plasma before $\sim 1\;\mev$, since they would 
give $\dneff \gtrsim 1$ during nucleosynthesis, which is clearly incompatible with the results of \App{dneffsec}.
In this section we derive the conditions for RH neutrino thermalization.

In the Standard Model, RH neutrinos (neutrinos with the ``wrong'' helicity) can be created in any interaction that produces LH neutrinos, i.e. weak interactions. 
The characteristic production rate of RH neutrinos in scattering reactions is 
$\Gamma_{\rm RH} \sim G_{\rm F}^2 T^5 (m_\nu/T)^2$~\cite{Dolgov:1994vu}. If weak interactions are the only interactions 
producing neutrinos, then RHNs do not thermalize as long as $m_\nu \lesssim \mathcal{O}(100\;\kev)$, which
is comfortably satisfied in the Standard Model. However, the interaction introduced in Eqs.~(1) and (2) of the main text is 
many orders of magnitude stronger than its weak counterpart, and the resulting RH neutrino production rate 
is much larger.

Thermalization is described by a Boltzmann equation of the form 
\beq
\dot \rho_{\nu_R} + 4H \rho_{\nu_R} = C,
\label{eq:boltzmann}
\eeq
where $\rho_{\nu_R}$ is the energy density in ``wrong helicity'' neutrinos and $C$ is the collision term encoding the 
processes that produce these neutrinos. Thermalization/equilibration occurs when $C \sim 4 H \rho_R^{\mathrm{eq}}$.
The dominant process responsible for RH neutrino production is the decay $\phi \rightarrow \nu \nu(\lambda=+1/2)$, where 
$\lambda = +1/2$ ($-1/2$) is a helicity label corresponding to RH (LH) neutrinos; 
scattering reactions such as $\nu\nu\rightarrow \nu\nu(\lambda)$ are suppressed 
by an additional factor of $g_\phi^2$. We therefore evaluate the 
collision term for $\phi\rightarrow \nu\nu(\lambda=+1/2)$:
\beq
C \approx 
2\int d\Phi_3 E_{\nu_R} |\mathcal{M}(\lambda=+1/2)|^2 f_\phi (1-f_\nu),
\label{eq:collision_term_decay1}
\eeq
where $d\Phi_3$ is the Lorentz-invariant phase-space (including the 
momentum conservation delta function) and $f_\phi$, $f_\nu$ are the phase-space distributions of 
$\phi$ and LH neutrinos. We have neglected the inverse decay contribution and the RH neutrino Pauli-blocking 
factors. These approximations are adequate for estimating the onset of equilibrium, assuming the 
initial abundance of RH neutrinos is negligible, i.e. $f_{\nu_R} \ll 1$. 
We have also multiplied the right-hand side by two to account for 
$\phi^*$ decays which can also produce wrong helicity neutrinos.

We use the Lagrangian in Eq.~(3) of the main text (with a complex $\phi$ for a Dirac neutrino) 
to evaluate the matrix element squared in the plasma frame, \emph{without} summing 
over one of the helicities~\cite{Dreiner:2008tw}
\begin{align}
|\mathcal{M}(\lambda)|^2 & \approx 
g_\phi^2\begin{cases}
\frac{E_\nu}{2E_{\nu_R}} (1 + \cos \theta)m^2_\nu & \lambda = + 1/2 \\ 
  m_\phi^2 & \lambda = - 1/2 
\end{cases},
\label{eq:ME_general_approx}
\end{align}
where $\theta$ is the angle between the $\nu_R$ and $\phi$ direction of motion, 
and we have kept only the leading terms in $m_\nu/m_\phi$.
Note that when $\cos\theta=-1$, the ``wrong''  helicity amplitude vanishes at this order 
as a result of angular momentum conservation (the next term is $\propto m_\nu^4$).
Using this result in \Eq{eq:collision_term_decay1}, we find that 
\beq
C  \approx \frac{g_\phi^2 m^2_\nu}{32\pi}\left(\frac{2\zeta(3) T^3}{\pi^2}\right) \mathcal{C}_+ (m_\phi/T),
\eeq
where the factor in parentheses is the number density of relativistic $\phi$ and 
the function $\mathcal{C}_+$ is $\sim 1$ for $m_\phi/T < 1$ and becomes Boltzmann-suppressed for $m_\phi/T > 1$.

We can apply the nucleosynthesis bound as computed in Section~\ref{dneffsec} if 
thermalization occurs before $T_{\rm dec} \sim 1\;\mev$. 
By using the approximate thermalization criterion below \Eq{eq:boltzmann},
we find that RH neutrinos thermalize before BBN if 
\beq
g_\phi \gtrsim 5\times 10^{-3} \left(\frac{\max(m_\phi, T_{\rm dec})}{\mev}\right)^{3/2}
\left(\frac{0.1\,\eV}{m_\nu}\right).
\eeq
This bound (evaluated using the full numerical $\mathcal{C}_+$) is shown in Figure 1 of the main text as a black dashed line.

\section{Supernova 1987A}
\label{snbound}

A new weakly coupled particle can change the behavior of the neutrino emission that was observed from the explosion of Supernova 1987A. The proto-neutron star cooling phase that was observed in large water Cherenkov detectors was qualitatively similar to the Standard Model-only expectation \cite{Burrows:1987zz}. If a new particle species $X$ carried away too much energy during the proto-neutron star cooling phase, the time over which neutrinos arrived  would have been unacceptably reduced \cite{Burrows:1988ah, Burrows:1990pk}. A semi-analytic criterion that the luminosity of this particle should obey is $L_X \leq L_\nu = 3\tenx{52} \erg/\s$ at times of order 1 second after the core bounce \cite{Raffelt:1996wa}. Following the procedure described in more detail in \cite{Chang:2016ntp}, we have 
\be \label{sn-lum}
\hspace{-0.1cm}L_\phi = \! \int_0^{R_\nu} \! \! \! dV \! \! \int \!\! \frac{d^3 k }{ (2\pi)^3 }
\,\omega \Gamma_\phi^{\rm prod} \exp \pL  \! -\int_r^{R_{\rm g}} \Gamma_\phi^{\rm abs} dr' \! \pR ,~~~~~
\ee
 where: the $\phi$ has four-momentum $(\omega, \vec k)$; $\Gamma_\phi^{\rm abs}$ is the $\phi$ absorptive width; $\Gamma_\phi^{\rm prod}$ is the $\phi$ production rate, which is related to the absorptive width in equilibrium by $ \Gamma_\phi^{\rm prod} = \exp( - \omega /T ) \Gamma_\phi^{\rm abs}$; $R_\nu$ is the radius of the neutrinosphere, outside of which neutrinos free-stream; and $R_{\rm g} = 100 \km$ is the radius inside of which neutrinos gain energy on average in elastic scattering events. We calculate $\Gamma_\phi$ including $\phi$ decay and $\phi$ annihilation to neutrino pairs, both of which are important for the masses of interest. 
 We have not included contributions from the neutrino effective potentials, which may be significant at small $m_\phi$~\cite{Brune:2018sab}. 
 We have also neglected neutrino Pauli blocking in $\Gamma_\phi$, which is important near the proto-neutron star core, since this will become unimportant between $R_\nu$ and $R_g$.

We find that $L_\phi$ given by \Eq{sn-lum} exceeds $L_\nu$ if $g_\phi$ is roughly in the range
\beq \label{sn87a-excl}
g_\phi^{\rm excl} \simeq \pL 5\tenx{-6} - 6\tenx{-5} \pR \times \frac1{ 1+m_\phi/\!\kev}.
\eeq
The sharp change in the shape of the bound at $m_\phi \simeq \kev$ is due to the fact that rate of decay and inverse decay, $\Gamma_{\phi \leftrightarrow \nu \nu} \propto g_\phi^2 m_\phi^2/T$, becomes subdominant to the annihilation rate, $\Gamma_{\phi \phi \leftrightarrow \nu \nu} \propto g_\phi^4 T$, for masses $m_\phi \lesssim g_\phi T_c$, where $T_c \sim \cO(30\mev)$ is the core temperature. These bounds are approximately compatible with those shown in \cite{Brune:2018sab} at masses above 10 keV. From \Eq{sn87a-excl}, we see that bounds arising from the luminosity of $\phi$ particles from Supernova 1987A are generally below the coupling range of interest in this work.

It is also interesting to understand the constraints on $g_\phi^{ee}$ arising from deleptonization of the core, which have been obtained in the $\sim \cO(\mev)$ mass range in \cite{Brune:2018sab} and which approximately overlap the range in \Eq{sn87a-excl}. At lower masses, this likely has an effect on the early phases of collapse and the collapse progenitor. Such a possibility was suggested in \cite{Kolb_1982} and was studied in the aftermath of Supernova 1987A by \cite{Fuller_1988}. This latter study found a constraint $g_\phi^{ee} < 3\tenx{-4}$, neglecting any $m_\phi$-dependence. Because these bounds are determined by physics at the beginning of the core collapse, when temperatures are $\sim \cO(\few\mev)$, the bound likely cuts off at $\sim 5 \mev$, similar to the $0\nu\beta\beta$ bounds cited above. A detailed study is of interest, but beyond the scope of this work.

\section{Strong Neutrino Self-Interactions via Sterile Neutrino Interactions}
\label{SecretNu}

Here we extend our exploration into the possibility that the interactions between the light neutrinos and the mediator $\phi$ are generated by mixing with sterile neutrinos, which we refer to as $N$. This requires interactions between $N$ and $\phi$ -- we refer to this coupling as $g_N$ -- and mixing between the light and sterile neutrinos, which we refer to as $|U|$. This scenario is appealing because the mixing $|U|$ may be temperature-dependent, allowing the constraints regarding thermalization of $\phi$ (and potentially $N$) prior to BBN. The mixing $|U|$ can then become large before $T \approx 100$ eV, allowing for strong interactions to modify CMB observations as desired by Ref.~\cite{Kreisch:2019yzn}. This could also explain the necessary $\Delta N_\mathrm{eff.} \approx 1$ at the time of CMB preferred by the fits in Ref.~\cite{Kreisch:2019yzn} if the increased mixing causes some particle to thermalize between BBN and CMB times.

If such a scenario exists, then the four-neutrino scattering with $\sigma \propto g_\phi^4$ now becomes $\sigma \propto g_N^4 |U|^8$ -- we identify the desired coupling $g_\phi \approx 10^{-1}$ with $g_N |U|^2$. Allowing $g_N$ to be as large as $\sqrt{4\pi}$, this dictates $|U|^2 \approx 10^{-2}$ by the time of CMB, $T \approx 100$ eV. In the main text and in Section~\ref{dneffsec}, we discussed the criteria for the thermalization of $\phi$ prior to BBN, contributing to $N_\mathrm{eff}$. In Section~\ref{dneffsec}, we focused on the values of $m_\phi$ that are incompatible with BBN observations. Additionally, there is the requirement that $g_\phi \gtrsim 10^{-11}$ in order for $\phi$ to thermalize. In order to avoid this constraint, allowing for a time-dependent mixing with $N$, we can enforce $|U|^2 \lesssim 10^{-11}$ until after BBN.

In the following, we explore the specifics of one scenario in which mixing can change rapidly between BBN and CMB, and we highlight the difficulties of realizing such a scenario. Ref.~\cite{Abazajian:2001nj} explored the temperature-dependent suppression of the sterile-active neutrino mixing, particularly depending on a lepton-number asymmetry to generate a large neutrino matter potential at early times in the universe. This temperature-dependent potential can lead to rapid changes in the sterile-active neutrino mixing, where $|U|^2 \propto T^{-8}$ for particular epochs, depending on parameters associated with the sterile neutrino. The suppression of $|U|^2$ is in effect when
\begin{equation}
\sqrt{2}G_F \eta_L T^3 \gg \frac{\Delta m^2}{T} \sqrt{1 - |U_0|^2},
\end{equation}
where $G_F$ is the Fermi constant, $\eta_L$ is the lepton number asymmetry, $\Delta m^2$ the new sterile-active neutrino mass-squared splitting. Here, we indicate the vacuum mixing as $|U_0|$ for clarity. In order to simultaneously avoid BBN constraints of thermalization of $\phi$ while still having an appreciable mixing during the time of the CMB, this suppression must be active when $T \approx 100$ keV - $1$ MeV, but no longer present by $T \approx 1$ keV. These two conditions place a constraint on $\Delta m^2/\eta_L$:
\begin{equation}
10^{-16}\ \mathrm{eV}^2 \lesssim \frac{\Delta m^2}{\eta_L} \lesssim 10^{-5}\ \mathrm{eV}^2.
\end{equation}
Even with order-one lepton number asymmetry, this indicates very small sterile-active neutrino mass-squared splittings. Ref.~\cite{Abazajian:2001nj} derives its results assuming there is one ``active'' neutrino species with which the sterile neutrino mixes. If $\Delta m^2 \lesssim 10^{-5}$ eV$^2$, the sterile neutrino is nearly degenerate with the active ones, and this formalism breaks down.

We conclude that, even if sterile neutrino interactions may be responsible for the large neutrino self interactions, further work is required to determine whether this mechanism can occur for the temperatures of interest. If, however, this mechanism may provide appreciable interactions at CMB times without thermalizing $\phi$ prior to BBN, we note that the results of Ref.~\cite{Kreisch:2019yzn} prefer large $N_\mathrm{eff}$ at the time of CMB. There is a possibility that the mixing $|U|^2$ turning on rapidly can cause $\phi$ to thermalize between BBN and CMB, providing the extra contribution to $N_\mathrm{eff}$ preferred by the fits in Ref.~\cite{Kreisch:2019yzn}.

\bibliography{nucmb}

\end{document}